\documentclass[twocolumn,twoside]{IEEEtran}
\usepackage{mathpple}
\usepackage{times}

\usepackage{amsmath}  
\usepackage{amssymb}  
\usepackage{mathrsfs} 

\usepackage{theorem}  
\usepackage{cite}     
\usepackage{comment}  

\usepackage{upref}
\usepackage{amsfonts}

\usepackage{verbatim}

\usepackage[dvipsnames,usenames]{color}
\usepackage{enumerate}

\usepackage{graphicx}
\usepackage{subfigure}

\usepackage{latexsym}

\newcommand{\code}{\ttfamily\bfseries}

\newcommand{\be}[1]{\begin{equation}\label{#1}}
\newcommand{\ee}{\end{equation}}

\newcommand{\bc}{\begin{center}}
\newcommand{\ec}{\end{center}}

\newcommand{\cC}{{\cal C}}

\newcommand{\cR}{{\cal R}}

\renewcommand{\leq}{\leqslant}

\newcommand{\Cref}[1]{Co\-rol\-la\-ry\,\ref{#1}}

\theoremstyle{plain} \theorembodyfont{\normalfont\slshape}

\newtheorem{thm}{Theorem$\!$}
\newenvironment{theorem}{\begin{thm}\hspace*{-1ex}{\bf.}}{\end{thm}}

\newtheorem{prop}[thm]{Proposition$\!$}

\newtheorem{lem}[thm]{Lemma$\!$}

\newtheorem{cor}[thm]{Corollary$\!$}
\newenvironment{corollary}{\begin{cor}\hspace*{-1ex}{\bf.}}{\end{cor}}

\newtheorem{prob}[thm]{Problem$\!$}

\newtheorem{defi}[thm]{Definition$\!$}
\newenvironment{definition}{\begin{defi}\hspace*{-1ex}{\bf.}}{\end{defi}}

\theorembodyfont{\normalfont}

\newtheorem{exam}{Example$\!$}

\newtheorem{remrk}{Remark$\!$}

\definecolor{Codecolor}{named}{White}  

\newcommand{\Copen}{\mbox{\{\kern-5.50pt\{}}
\newcommand{\Cclose}{\mbox{\}\kern-5.50pt\}}}
\newcommand{\Cslash}{\mbox{$\backslash\kern-6.02pt\backslash$}}

\begin{document}
\title{\textbf{\huge{When Do WOM Codes Improve the Erasure Factor in Flash Memories?}}}

\author{\textbf{Eitan Yaakobi},\! \IEEEauthorblockN{\textbf{Alexander Yucovich},\! \textbf{Gal Maor},\! and \textbf{Gala Yadgar}}

\IEEEauthorblockA{Computer Science Department, Technion -- Israel Institute of Technology, Haifa 32000, Israel\\}
{\code  \{yaakobi,galmaor,gala\}@cs.technion.ac.il, yucovich@campus.technion.ac.il}\,
\vspace{-5ex}}
\maketitle
\thispagestyle{empty}
\pagestyle{empty}

\begin{abstract}
Flash memory is a \emph{write-once} medium in which reprogramming cells requires first erasing the block that contains them. The lifetime of the flash is a function of the number of block erasures and can be as small as several thousands. To reduce the number of block erasures, pages, which are the smallest write unit, are rewritten \emph{out-of-place} in the memory. A \emph{Write-once memory (WOM) code} is a coding scheme which enables to write multiple times to the block before an erasure. However, these codes come with significant rate loss. For example, the rate for writing twice (with the same rate) is at most 0.77.

In this paper, we study WOM codes and their tradeoff between rate loss and reduction in the number of block erasures, when pages are written uniformly at random. First, we introduce a new measure, called \emph{erasure factor}, that reflects both the number of block erasures and the amount of data that can be written on each block. A key point in our analysis is that this tradeoff depends upon the specific implementation of WOM codes in the memory. We consider two systems that use WOM codes; a conventional scheme that was commonly used, and a new recent design that preserves the overall storage capacity. While the first system can improve the erasure factor only when the storage rate is at most 0.6442, we show that the second scheme always improves this figure of merit.
\end{abstract}
\vspace{-3ex}

\section{Introduction}\label{sec:intro}
\renewcommand{\baselinestretch}{0.943}\normalsize\noindent
Flash memories are, by far, the most important type of non-volatile memory (NVM) in use today. Flash devices are employed widely in mobile, embedded, and mass-storage applications, and the growth in this sector continues at a staggering pace. The most conspicuous property of flash-storage technology is its inherent asymmetry between cell programming and cell erasing. While it is fast and simple to increase a cell level, reducing its level requires a long and cumbersome operation of first erasing the entire block that contains it and only then programming the cell. Such block erasures are not only time-consuming, but also degrade the lifetime of the memory, which can typically tolerate $10^3-10^5$ block erasures. Therefore, finding algorithms for increasing its lifetime despite this asymmetric programming behavior has become an important challenge.

A flash memory chip is built from floating-gate cells. A group of cells constitute a page, which is the smallest write unit, and the pages are organized in blocks, which are the smallest erase unit. Since pages can be updated only if their accommodating block is first erased, write update requests are performed \emph{out-of-place}. Thus, when a page is updated, its previous location is marked as \emph{invalid}. In order to accommodate this write procedure, the amount of physical storage has to be larger than the available logical storage. The ratio between the size of additional storage and logical storage is called \emph{over-provisioning}. Furthermore, whenever there are no available blocks to accommodate page write requests, \emph{garbage collection (GC)} is invoked to clean, i.e. erase, blocks for additional page writes. However, when a block is chosen to be cleaned by GC, its valid pages must be read and rewritten to a clean block, thereby increasing the total number of pages written to the memory. The \emph{write amplification} is the ratio between the number of physical page writes and the number of logical page writes. 

Reducing the write amplification is crucial as it directly affects the memory performance and its lifetime. In general, there is a direct relation between over-provisioning and write amplification. Increasing over-provisioning reduces the write amplification~\cite{D14}. However, high over-provisioning means that a large area of the memory is not exploited to store information. Thus, understanding the connection between the two measures is very important for optimizing the design of flash memories.

Write-once memory (WOM) codes were first introduced in 1982 by Rivest and Shamir~\cite{RS82}. In the binary version, write-once memory cells can only be irreversibly programmed from a value of zero to a value of one. The motivation to study WOM came from storage media like punch cards and optical disks. A renewed interest in WOM codes came along in the past years as a result of the tremendous research work on coding for flash memories. Flash memories impose similar constraints in which the level of each cell can only increase, and can be decreased only if its entire block is first erased. Thus, a WOM-code can be applied in flash memories to enable additional writes without first having to erase the block. WOM codes in flash memories were investigated both theoretically with respect to the number of block erasures, and practically by simulations; see e.g.~\cite{BB13, G_etal09, JCS13, LKY12, OC14,Z_etal13}.

The reduction in the number of block erasures via WOM codes is beneficial for extending device lifetime. However, this benefit comes with a significant price of non-negligible increase in the redundancy, thereby decreasing the over-provisioning in the memory. Note that when using WOM codes, it is possible to write more pages with less erasures since every block can be written more than once. Therefore, write amplification is not the right figure of merit for this method. Thus, we introduce a new measure, called \emph{erasure factor}, which is the ratio between the number of block erasures and number of logical block write requests. When WOM codes are not used, the erasure factor is equivalent to the write amplification.

The main goal of this paper is to analyze the erasure factor when using WOM codes, while pages are written uniformly at random. In order to have a fair comparison with systems that do not use WOM codes, we fix the over-provisioning and then compare the erasure factor. We analyze two different implementations of WOM codes. Conventional implementations of WOM codes used the codes in the page level such that each page is written multiple times, and hence incorporated capacity loss. We compare these with a new approach, recently proposed in~\cite{YYS15}, that avoids the capacity loss by encoding information into more than a single page after the first write. 

The rest of the paper is organized as follows. In Section~\ref{sec:def}, we introduce the necessary background on flash memories and WOM codes, and formally define the problem we study in the paper. In Section~\ref{sec:no_WOM}, we review and state the results on the connection between over-provisioning and the erasure factor without using WOM codes. In Section~\ref{sec:base_WOM}, we study the conventional implementation of WOM codes and similarly analyze this connection. We then continue in Section~\ref{sec:CP-WOM} to study a more efficient implementation of WOM codes in which no rate loss is incurred and similarly study its erasure factor.
\vspace{-2ex}
\section{Definitions and Problem Statement}\label{sec:def}
\subsection{Flash Memory Structure}
Flash memories consist of floating-gate cells that can typically store a single bit, two bits, or three bits. The cells are organized into blocks which usually contain 64-384 pages, where the size of a page ranges between 2KB and 16KB~\cite{G_etal09}. Due to the inherent asymmetry between programming and erasing, flash memories perform page writes \emph{out-of-place}. This write procedure introduces the following concepts:
\begin{itemize}
\item \emph{\textbf{Flash Translation Layer (FTL)}:} The FTL is responsible for mapping logical locations to physical ones.
\item \emph{\textbf{Over-provisioning (OP)}:} The ratio between the amounts of additional storage and logical storage. This overhead is necessary to accommodate out-of-place writes.
\item \emph{\textbf{Garbage Collection (GC)}:} The process in charge of cleaning blocks in order to free more space for writing.
\item \emph{\textbf{Write Amplification (WA)}:} The ratio between the number of physical page writes and the number of logical page write requests.
\end{itemize}

The following summarizes the setup, notations, and assumptions we use throughout the paper. These notations hold for a flash memory device, e.g. a solid state drive.
\begin{enumerate}
\item Every block has $Z$ pages, each of size $s$KB. There are $T$ physical pages and $U$ logical pages, where both $T$ and $U$ are a multiple of $Z$.
\item The over-provisioning is $\rho=(T-U)/U$ and $\alpha = U/T = 1/(\rho+1)$ is the \emph{storage rate}, which is the ratio between logical data and physical storage.
\item $WA=P/L$, where $L$ is the number of write requests of logical pages and $P$ is the number of resulting physical page writes. We also define $L/Z$ to be the number of \emph{logical block writes}.\vspace{-1ex}
\end{enumerate}

\vspace{-2ex}
\subsection{WOM Codes}
WOM codes were first introduced by Rivest and Shamir in 1982~\cite{RS82}, and were found to be very relevant in the context of rewriting algorithms for flash memories. In this setup, the memory consists of $n$ cells and the goal is to maximize the number of bits which can be written to the memory in $t$ writes, while guaranteeing that each cell is changed only from 0 to 1. The most famous example of a WOM code is the one given by Rivest and Shamir for writing two bits twice using only three cells~\cite{RS82}. In their work, they also analyzed the bounds on the amount of information that can be stored in a WOM. Since then, more constructions were given in the 1980's and 1990's, e.g.,~\cite{CGM86} as well as capacity analysis, e.g.,~\cite{FH99,H85}. Several more constructions of WOM codes were recently given; see e.g.~\cite{BS13,S12A,YKSVW12}

Assume $t$ messages are written to the memory, consisting of $n$ cells. On the $i$-th write, $1\leq i\leq t$, the message size is $M_i$. The \emph{rate} on the $i$-th write is defined to be $\cR_i =\frac{\log_2 M_i}{n}$, and the \emph{sum-rate} is $\cR_{\textmd{sum}} = \sum_{i=1}^{t}\cR_i$. We consider two types of WOM codes~\cite{YKSVW12}. In a \emph{fixed-rate} WOM code, the rate on all writes is the same, while in a \emph{variable-rate WOM code} the rate may vary on each write.
The \emph{capacity region} of a $t$-write WOM is the set of all achievable rate tuples. For the binary case, the capacity region was found in~\cite{FH99,H85,RS82}. It was also proved that the maximum achievable sum-rate for a WOM code with $t$ writes is $\log_2(t+1)$. Similar results were given for fixed-rate WOM codes~\cite{H85}. For example the maximum sum-rate of a two-write fixed-rate WOM code is 1.54.

\vspace{-2ex}
\subsection{Problem Setup}
The main goal of this work is to study the connection between the over-provisioning $\rho$ (or storage rate $\alpha$) and the number of block erasures. This connection depends upon the over-provisioning value, GC algorithm, and the probability distribution of the page write requests\footnote{and also on $Z$ but we assume in the paper that $Z$ is large enough to avoid this dependency.}. We assume in this work that requests are uniformly distributed over the $U$ logical pages. We follow the observation from~\cite{HH10} claiming that \emph{greedy GC} is optimal for uniform distribution, where greedy GC always chooses the block with the minimum number of valid pages for cleaning. We also assume that greedy garbage collection is invoked whenever there are no more clean blocks. That is, we don't require a minimum fraction of available blocks since the analysis is very similar to the one without this requirement~\cite{D14}. 

WOM codes allow to write the blocks multiple times before an erasure. Thus, WA is not the right figure of merit since it is possible to write more pages and yet erase less. Hence, we introduce a new measure that better characterizes this behavior.\vspace{-2ex}
\begin{definition}
The \textbf{erasure factor} $EF$ in a flash memory system is the ratio between the number of block erasures $E$ and the number of logical block writes $L/Z$. That is,
\vspace{-1ex}$$EF=\frac{E}{L/Z}\vspace{-1ex}.$$
\end{definition}

Note that if no rewriting code is used then $EF=WA$.
In the rest of the paper, we study the erasure factor of several systems with and without WOM codes, demonstrating how the specific usage of WOM codes directly affects this figure of merit.

\section{The Relation between Over-provisioning and Erasure Factor}\label{sec:no_WOM}

The relation between the write amplification and over-provisioning has received a significant attention in recent years due to its importance to the lifetime of flash memories, see e.g.~\cite{D14,HH10,SA13}. Of the numerous works in this area, we consider two recent studies which we believe give an accurate model of this analysis~\cite{D14,SA13}. The proof given here is based upon the analysis in these two studies and we give it in completeness since its understanding is crucial to the results in the paper.
For the purpose of our discussion, we call the system in these studies, that does not use WOM codes, the \emph{\textbf{baseline system}}.
\vspace{-2ex}
\begin{theorem}\label{th:base}
The number of block erasures $E$ and the erasure factor $EF_1(\alpha)$ of the baseline system are given by\vspace{-1ex}
\begin{equation}
E = \frac{P}{L} = \frac{L}{Z(1-\alpha')},  \ \ \ EF_1(\alpha) = \frac{1}{1-\alpha'},\vspace{-1ex}
\end{equation}
where $\alpha = \frac{\alpha'-1}{\ln(\alpha')}$, or $\alpha' = -\alpha\cdot W\left(  -\frac{1}{\alpha}e^{-1/\alpha} \right)$, and $W(x)$ is the Lambert $W$ function.
\end{theorem}
\begin{IEEEproof}
For $0\leq i\leq Z$, let $N(i)$ be a random variable corresponding to the number of blocks with $i$ valid pages, so $\sum_{i=0}^{Z}N(i)=T/Z.$
If we denote by $Y$ the expected number of valid pages when a block is erased, then for $0\leq i\leq Y-1$, $N(i)=0$, and $N(Y)$ is relatively small enough.  
We assume that the system is in steady state and thus the expected value of $N(i)$ doesn't change over time\footnote{These properties are taken from~\cite{BI10,D14} where this process is modeled as a Markov chain and the number of blocks with a given number of valid pages is fixed for analysis purposes.}. According to this assumption, we also get that for $Y+1\leq i\leq Z$,
$$iN(i) = C,$$ for some constant $C$, or $N(i) = (Y+1)N(Y+1)/i$.
Therefore, we get\footnote{While there are better approximations to the differences between two Harmonic series, we choose this one since it provides better expressions which can be analyzed without dependency on the number of pages in a block.}
\vspace{-1ex}
\begin{align*}
T/Z &= \sum_{i=0}^{Z}N(i)=\sum_{i=Y+1}^{Z}N(i) = \sum_{i=Y+1}^{Z}(Y+1)N(Y+1)/i & \vspace{-2ex}\\ \vspace{-2ex}
  & = (Y+1)N(Y+1)\sum_{i=Y+1}^Z\frac{1}{i}& \\
  & \approx (Y+1)N(Y+1)(\ln(Z)-\ln(Y)) & \\
  & = (Y+1)N(Y+1)\ln(Z/Y). &\vspace{-1ex}
\end{align*}
We also have that\vspace{-1ex}
$$U = \sum_{i=0}^ZiN(i) = \sum_{i=Y+1}^ZiN(i) = (Z-Y)(Y+1)N(Y+1).\vspace{-1ex}$$
Together, we get that\vspace{-1ex}
$$(Y+1)N(Y+1) = \frac{T/Z}{\ln(Z/Y)}= \frac{U}{Z-Y},\vspace{-1ex}$$
or\vspace{-1ex}
$$\alpha = \frac{U}{T}= \frac{Z-Y}{Z\ln(Z/Y)} = \frac{Y/Z-1}{\ln(Y/Z)} = \frac{\alpha'-1}{\ln(\alpha')}.\vspace{-1ex}$$
where $\alpha' = Y/Z$, and is given by $\alpha'=-\alpha\cdot W\left(  -\frac{1}{\alpha}e^{-1/\alpha} \right)$.

Now, we deduce that for every $Z-Y$ logical page writes, $Z$ physical pages are written. Hence, $P=L\cdot \frac{Z}{Z-Y} = \frac{L}{1-\alpha'}$, and
\vspace{-1ex}
$$ E = \frac{P}{Z} = \frac{L}{Z(1-\alpha')}, EF_1(\alpha) = \frac{E}{L/Z} = \frac{1}{1-\alpha'}.\vspace{-3ex}$$
\end{IEEEproof}
\vspace{-3ex}
\section{Analysis of the Naive-WOM System}\label{sec:base_WOM}

In this section, we take a first step in analyzing the erasure factor when using WOM codes. However, as we shall later see, this analysis depends on the specific implementation of WOM codes in the memory. In order to have a fair comparison with the baseline system, we carry out this comparison while fixing $\alpha$, the ratio between logical and physical storage. Furthermore, in order to minimize the modifications in the architecture, we assume that the block size is fixed, however we will allow to change the size of the physical pages and accordingly the number of physical pages in a block.

Let us start by describing the conventional setup to implement WOM codes in flash. This setup was tested experimentally in several studies, see e.g.~\cite{G_etal09, JCS13} along with analytical derivations in~\cite{LKY12}.
Assume a two-write fixed-rate WOM code is used with individual rate $R$ on each write, so $R\leq 0.77$~\cite{H85}.
Here, WOM codes allow to write each block twice before an erasure. First, all pages are written sequentially in the block and after the block is chosen by GC, it is possible to write to the invalid pages. Thus, every block can be either on a first or second write. The modifications in the system setup compared to the baseline system are summarized as follows:
\begin{enumerate}
\item A two-write WOM code with fixed-rate $R$ on each write is used to write all pages. Thus, the size of every physical page is $\frac{1}{R}sKB$ so it can accommodate a write of a logical page which is encoded by the WOM code's encoder.
\item As a result of increasing the physical page size, the number of pages in a block reduces to $Z' = RZ$, so the block size remains the same. Accordingly, the total number of physical pages is also reduced to $T'=RT$. Hence, the ratio $\beta$ between the number of logical pages and physical pages is \vspace{-1ex}
$$\beta = \frac{U}{T'} = \frac{U}{RT} = \frac{\alpha}{R},\vspace{-1ex}$$
and thus the storage rate is $\alpha=\beta R$.
\item We use the same greedy GC as in the baseline system, and hence the block with the minimum number of valid pages is chosen by the GC. If the block is on first write, then it is moved to second write, and since it is not erased, its valid pages remain in the block. If the block is on second write, then its valid pages are rewritten on an available block and the block is erased.
\end{enumerate}

We call this method of implementing WOM codes the \emph{\textbf{naive-WOM system}}. Since the value of $\beta$ is at most 1, we can compare the baseline system and the naive-WOM system only for $\alpha \leq R=0.77$. The next theorem states the result of the erasure factor under this setup. The proof is omitted due to the lack of space and since it is a direct application of Theorem~\ref{th:base}.
\vspace{-4ex}
\begin{theorem}\label{th:base_WOM}
For any $\alpha \leq R$, under uniform writing with greedy GC, the erasure factor $EF_2(\alpha)$ of the naive-WOM system is given by\vspace{-2ex}
$$EF_2(\alpha) = \frac{1}{2(1-\beta')}\vspace{-1ex}$$ where $\beta' = -\beta\cdot W\left(  -\frac{1}{\beta}e^{-1/\beta} \right)$, and $\beta = \frac{\alpha}{R}$.
\end{theorem}\vspace{-1ex}

Finally, we can compare these two systems and find the values of $\alpha$ in which the naive-WOM system is superior to the baseline system with respect to the erasure factor.\vspace{-1ex}
\begin{corollary}
The naive-WOM system for $t=2$ with rate $R$ has better erasure factor than the baseline system if \vspace{-1ex}
\begin{equation} \hspace{-0.5ex}{1+\alpha\cdot W\left(  -\frac{1}{\alpha}e^{-1/\alpha} \right)} \leq {2\left(1+\frac{\alpha}{R}\cdot W\left(  -\frac{R}{\alpha}e^{-\frac{R}{\alpha}} \right)\right)}.\vspace{-1ex}
\end{equation}
In particular, for $R=0.77$, it has better erasure factor for $\alpha\leq 0.6442$.
\end{corollary}

\begin{figure}%
\centering
\includegraphics[width =0.6\linewidth]{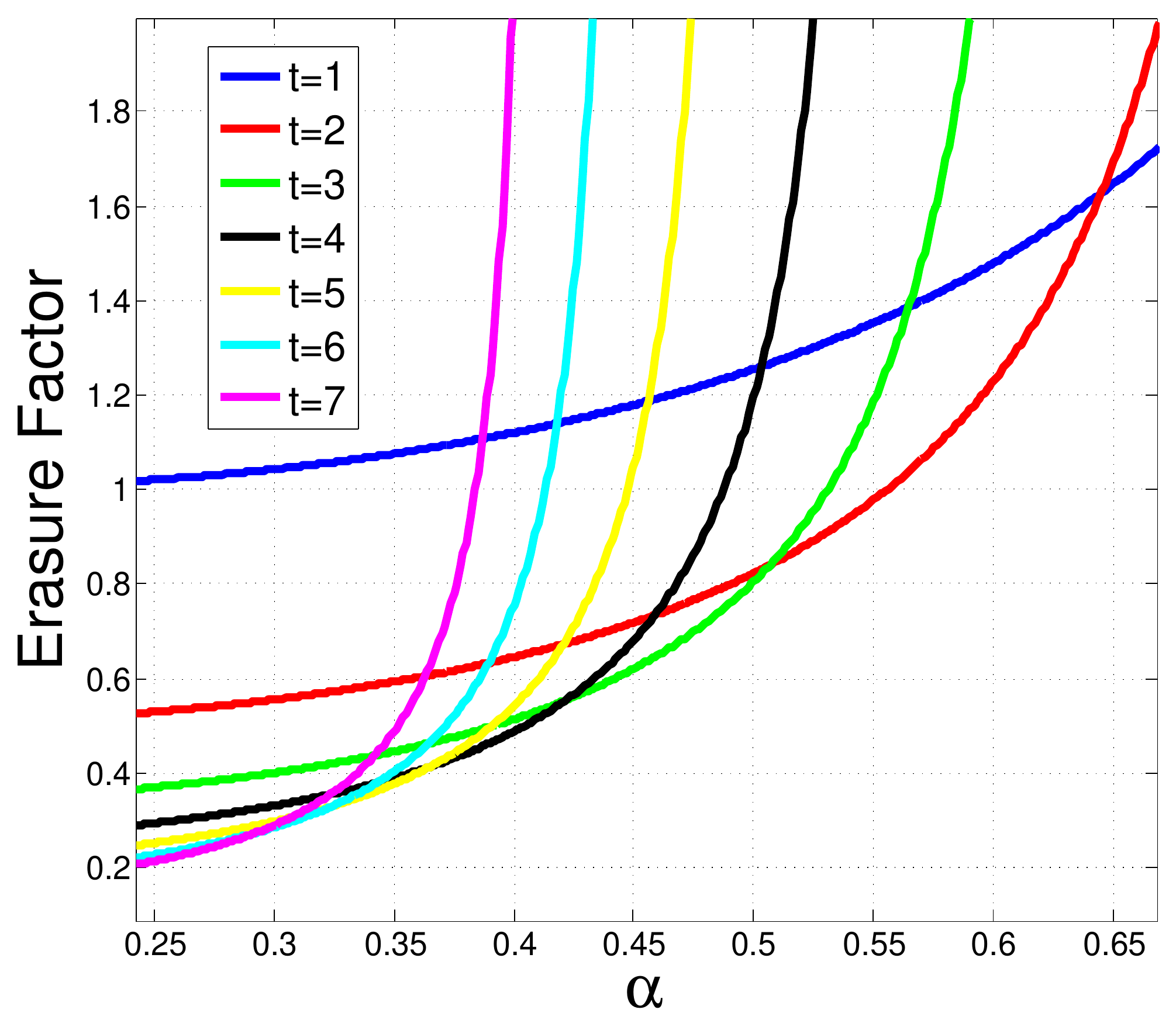}%
\vspace{-2ex}\caption{A comparison between the erasure factors of the baseline and naive-WOM systems with multiple writes.}
\label{fig:WA1}%
\vspace{-3ex}
\end{figure}

The extension to multiple writes is immediate. Assume the individual rate on each write is $R_t$ where the optimal values of $R_t$ are given in~\cite{H85}. Then, as before, we have that the erasure factor $EF_{t}(\alpha)$ is given by \vspace{-1ex}
\begin{equation}EF_{t}(\alpha) =  \frac{1}{t\left(1+\frac{\alpha}{R_t}\cdot W\left(  -\frac{R_t}{\alpha}e^{-\frac{R_t}{\alpha}} \right)\right)}.\vspace{-1ex}\end{equation}
In Fig.~\ref{fig:WA1}, we plot the curves of $EF_{t}(\alpha)$ $\alpha$ for $2\leq t\leq 7$, and the respective optimal $R_t$ values.

\vspace{-2ex}
\section{Analysis of the Capacity-Preserving-WOM System}\label{sec:CP-WOM}
The main disadvantage of the naive-WOM system is its high rate penalty and thus an increase in the over-provisioning. In fact, this is a major caveat which prevented a wide adoption of WOM codes in flash memory devices. Furthermore, it also requires to either change the page or block size, which introduces another level of complication in the architecture design.

These two disadvantages were recently resolved in~\cite{YYS15} by proposing an implementation of WOM codes which requires neither storage rate loss nor changing the page size. The full details of this implementation include considerations of performance, parallelism and the complexities of the encoder and decoder~\cite{YYS15}. For the sake of analyzing the erasure factor, we take a less restrictive approach which does not affect the analysis results. We call this scheme the \emph{\textbf{Capacity-Preserving WOM system}} or in short \emph{\textbf{CP-WOM system}}. We only give the main ideas of this implementation as the full details appear in~\cite{YYS15}.\\
\emph{\textbf{Choice of a WOM code}:} The capacity region of a two-write WOM is given by the formula~\cite{FH99,H85}
\vspace{-1ex}$$
\cC_{2} = \{(R_1,R_2)\ |\ \exists p\in [0,0.5], R_1\leq h(p),R_2\leq 1-p\},
\vspace{-1ex}$$
where $h$ is the binary entropy function. In particular, the rate tuple $(1,0.5)$ belongs to this region, and thus we assume that a two-write WOM code with these rates exists. In practice, we note that it is hard to find such codes that are successful in the worst case. Thus, the work in~\cite{YYS15} used \emph{polar WOM codes} which are capacity achieving and their success is guaranteed with high probability\footnote{Note that it is possible to retry the encoding of polar WOM codes with a different dither, thereby the failure probability is negligible for any practical purpose.}. It is also possible to use the recent design of capacity achieving WOM codes by LDGM codes~\cite{EHLB15}. Note that since we use WOM codes with rate $(1,0.5)$ there is no capacity loss on the first write and thus the storage rate remains the same. \\
\emph{\textbf{FTL Consideration}:} From the FTL point of view, every block can be in one of two states. When it is clean, pages are written to it sequentially such that there is no capacity loss (the rate of the WOM code on this first write is 1). After a block is chosen by the GC, it is not physically erased. Instead, its valid pages reside in the block and the invalid pages can be used to accommodate a second write. Now, every logical page will be encoded and written into two physical pages in the block (since the rate of the second write is 0.5). When this block is chosen again by the GC, its valid pages are copied and rewritten in order to allow a physical erasure of the block.\\ 
\emph{\textbf{Greedy GC Policy}:} The blocks in this implementation can be in two different states: first or second write.
The greedy GC policy is characterized by a parameter $0\leq \gamma_1\leq 1$. Let $B_1,B_2$ be the blocks with the minimum number of valid logical pages on a first, second write, respectively. If the number of valid pages in $B_1$ is at most $Y_1=\gamma_1\cdot Z$ then this block is moved from first to second write, and otherwise the block $B_2$ is physically erased and its valid pages are copied and written to an available block. 
The value of $\gamma_1$ in the greedy GC which optimizes the number of erasures and the corresponding erasure factor is found in the next theorem.\vspace{-2ex}
\begin{theorem}\label{th:CP_WOM}
For any storage ratio $\alpha$ and greedy GC with parameter $\gamma_1$, the erasure factor of the CP-WOM system is given by \vspace{-2ex}
\begin{equation}EF_2'(\alpha, \gamma_1)=  \frac{1}{3/2-\gamma_1/2-\gamma_2},\vspace{-1ex}\end{equation}
where $\gamma_2$ satisfies the relation\vspace{-1ex}
\begin{equation}\gamma_2 = -\alpha W\left(  -\frac{1}{\alpha} e^{\ln\left(\frac{1+\gamma_1}{2\gamma_1}\right) + \frac{\gamma_1-3}{2\alpha}}\right).\vspace{-1ex}\end{equation}
The optimal erasure factor is given by \vspace{-1ex}
\begin{equation}EF^*_2(\alpha)=  \min_{0\leq \gamma_1 \leq 1}\left\{ EF_2'\left(\alpha, \gamma_1\right)   \right\}. \vspace{-1ex}\end{equation}
\end{theorem}
\begin{IEEEproof}
Let $Y_1=\gamma_1\cdot Z$ and let us denote by $Y_2$ the expected number of valid pages when a block on a second write is physically erased and let $\gamma_2=Y_2/Z$. We will determine the relation between the values of $Y_1$ and $Y_2$. For $0\leq i\leq Z$, we denote by $N_1(i), N_2(i)$ the number of blocks with $i$ valid pages on a first, second write, respectively. Notice first that for $i\leq  Y_1$, $N_1(i) = 0$ and for $i\leq Y_2$, $N_2(i) = 0$. Furthermore, when a block is moved from first to second write, it already contains $Y_1$ valid pages. Since every logical page is written into 2 available pages, the total number of logical pages this block can accommodate is at most $Y_1+(Z-Y_1)/2 = (Z+Y_1)/2$ and thus $N_2(i) = 0$ for $i>(Z+Y_1)/2$.

We follow the same steps of the proof of Theorem~\ref{th:base} to have the following equations:
\begin{align*}
& (Y_1+1)N_1(Y_1+1)= \cdots = ZN_1(Z) & \\
=& (Y_2+1)N_2(Y_2+1) = \cdots = \frac{Z+Y_1}{2}N_2(\frac{Z+Y_1}{2}). &
\end{align*}
According to these definitions, a block can accommodate $Z$ page writes on the first write and $(Z-Y_1)/2$ more page writes on the second write. Furthermore, on every block erasure, $Y_2$ pages are rewritten, so the number of erasures is given by\vspace{-1ex}
$$E=\frac{L+EY_2}{Z+(Z-Y_1)/2}\vspace{-1ex}$$
or\vspace{-1ex}
$$E = \frac{L}{3Z/2-Y_2-Y_1/2} = \frac{L}{Z}\cdot \frac{1}{3/2-\gamma_1/2-\gamma_2}.\vspace{-1ex}$$
Hence, the erasure factor, as a function of both $\gamma_1$ and $\gamma_2$ is $1/(3/2-\gamma_1/2-\gamma_2)$.

Following the rest of the steps from Theorem~\ref{th:base} we get
\begin{small}
\begin{align*}
T/Z &= \sum_{i=0}^{Z}(N_1(i)+N_2(i))  =\sum_{i=Y_1+1}^{Z}N_1(i) + \sum_{i=Y_2+1}^{\frac{Z+Y_1}{2}}N_2(i)  & \\
& = \hspace{-1.5ex}\sum_{i=Y_1+1}^{Z}\hspace{-1.5ex}\frac{(Y_1+1)N_1(Y_1+1)}{i}  +\hspace{-1.5ex} \sum_{i=Y_2+1}^{\frac{Z+Y_1}{2}}\hspace{-1.5ex}\frac{(Y_1+1)N_1(Y_1+1)}{i} & \\
& = (Y_1+1)N_1(Y_1+1)\left( \sum_{i=Y_1+1}^Z\frac{1}{i} +\sum_{i=Y_2+1}^{\frac{Z+Y_1}{2}}\frac{1}{i}  \right)& \\
& \approx (Y_1+1)N_1(Y_1+1)\left(\ln\left(\frac{Z}{Y_1}\right)+\ln\left(\frac{Z+Y_1}{2Y_2}\right)\right) & \\
& = (Y_1+1)N_1(Y_1+1)\ln\left(\frac{1+\gamma_1}{2\gamma_1\gamma_2}\right). &
\end{align*}
\end{small}
As before, we also have
\begin{align*}
U & = \sum_{i=0}^ZiN_1(i)  + \sum_{i=0}^ZiN_2(i) & \\
   & = \sum_{i=Y_1+1}^ZiN_1(i) + \sum_{i=Y_2+1}^{\frac{Z+Y_1}{2}}iN_2(i)  & \\
   & = (3Z/2-Y_1/2-Y_2) (Y_1+1)N_1(Y_1+1). &
\end{align*}
Thus we get \vspace{-1ex}
$$(Y_1+1)N_1(Y_1+1) = \frac{U}{3Z/2-Y_1/2-Y_2} =  \frac{T/Z}{\ln\left(\frac{1+\gamma_1}{2\gamma_1\gamma_2}\right)},\vspace{-1ex}$$
or \vspace{-1ex}
$$ \alpha = \frac{3/2-\gamma_1/2-\gamma_2}{\ln\left(\frac{1+\gamma_1}{2\gamma_1\gamma_2}\right)},\vspace{-1ex}$$
that is\vspace{-1ex}
\begin{equation}\label{eq:gamma2}
\gamma_2 = -\alpha W\left(  -\frac{1}{\alpha} e^{\ln\left(\frac{1+\gamma_1}{2\gamma_1}\right) + \frac{\gamma_1-3}{2\alpha}}\right).
\vspace{-1ex}
\end{equation}
Hence, the erasure factor, as a function of $\alpha$ and $\gamma_1$ is
$$EF_2'(\alpha,\gamma_1) = \frac{1}{3/2-\gamma_1/2-\gamma_2},$$
where $\gamma_2$ is given by~(\ref{eq:gamma2}). Lastly, since we can choose the threshold $\gamma_1$, the value $EF^*_2(\alpha)$ is achieved by minimizing the value of $EF_2'(\alpha, \gamma_1)$ under the condition in~(\ref{eq:gamma2}).
\end{IEEEproof}

\begin{figure}%
\centering
\includegraphics[width =0.6\linewidth]{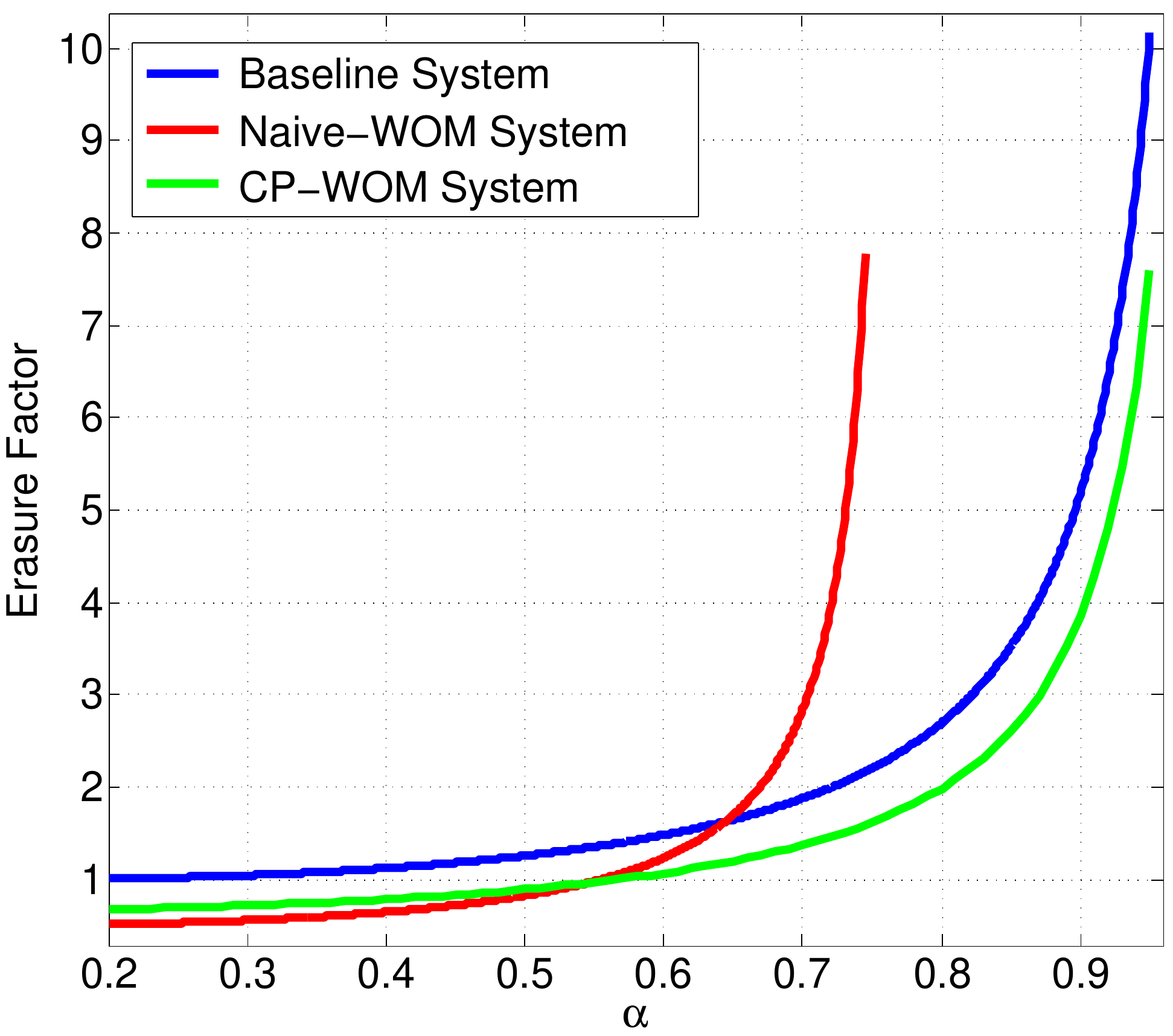}%
\vspace{-2ex}\caption{A comparison between the erasure factors of different systems.}
\label{fig:WA2}%
\vspace{-3ex}
\end{figure}

In Fig.~\ref{fig:WA2}, we plot the curves of the erasure factors of the baseline, naive-WOM, and CP-WOM systems for two writes. We see that the CP-WOM system always has a better erasure factor than the baseline system. The naive-WOM system has better erasure factor than the CP-WOM system roughly for $\alpha\leq 0.54$. This is not surprising since for small values of $\alpha$, no pages are copied by the GC, so the best EF of the CP-WOM system is $1/1.5=2/3$ 
while the best  $EF$ of the naive-system is $1/2$. The extension to multiple writes follows similar steps.

\vspace{-1.5ex}
\begin{theorem}\label{th:CP-WOM}
For any storage ratio $\alpha$ and a CP-WOM system for $t$ writes, the erasure factor is given by
\begin{equation}EF^*_t(\alpha)=  \min_{0\leq \gamma_1, \gamma_2,\ldots, \gamma_{t -1}\leq 1}\left\{ EF_t'\left(\alpha, \gamma_1,\gamma_2,\ldots, \gamma_{t-1}\right)   \right\}, \end{equation}
where\vspace{-1ex}
\begin{equation} EF_t'(\alpha,\gamma_1,\gamma_2,\ldots,\gamma_{t-1}) = \frac{1}{2-\frac{1}{2^{t-1}}- \frac{\sum_{j=1}^{t-1}\gamma_j}{2}-\gamma_t},\vspace{-1ex}\end{equation}
and $\gamma_t$ satisfies\vspace{-2ex}
\begin{equation}\alpha = \frac{2-\frac{1}{2^{t-1}}- \frac{\sum_{j=1}^{t-1}\gamma_j}{2}-\gamma_t}{\sum_{j=1}^{t}\ln\left(\frac{1+2^{j-2}\gamma_{j-1}}{2^{j-1}\gamma_j}\right)},\vspace{-1ex}\end{equation}
while $\gamma_0=0$.
\end{theorem}

\section{Conclusion}\label{sec:conc}
The primary goal of this paper is to establish the foundations in analyzing the erasure factor (as an alternative measure to the unsuitable WA) when using WOM codes. This analysis is very important since it answers the fundamental questions regarding the potential benefit of using WOM codes in flash memories and similar memories. We hope that similar studies will be conducted in order to analyze the performance of other coding schemes with respect to the erasure factor. Some future research directions are summarized below.
\begin{enumerate}
\item A tighter approximation of the erasure factor when blocks have a relatively small number of pages.
\item Analysis of other systems implementing WOM codes and additional coding schemes.
\item Analysis of non-binary WOM codes.
\item Analysis of workloads other than the uniform writing; see for example~\cite{D14,SA13}.\vspace{-1ex}
\end{enumerate}

\section*{Acknowledgment}
This was supported in part by the Israel Science Foundation (ISF) Grant No. 1624/14.	

\end{document}